# Coordinated appendages accumulate more energy to self-right on the ground

Qihan Xuan[1], Chen Li[1]

***Abstract*—Animals and robots must right themselves after flipping over on the ground. The discoid cockroach pushes its wings against the ground in an attempt to dynamically self-right by a somersault. However, because this maneuver is strenuous, the animal often fails to overcome the potential energy barrier and makes continual attempts. In this process, the animal flails its legs, whose lateral perturbation eventually leads it to roll to the side to self-right. Our previous work developed a cockroach-inspired robot capable of leg-assisted, winged self-righting, and a robot simulation study revealed that the outcome of this strategy depends sensitively on wing-leg coordination (measured by the phase between their motions). Here, we further elucidate why this is the case by developing a template to model the complex hybrid dynamics resulting from discontinuous contact and actuation. We used the template to calculate the potential energy barrier that the body must overcome to self-right, mechanical energy contribution by wing pushing and leg flailing, and mechanical energy dissipation due to wing-ground collision. The template revealed that wing-leg coordination (phase) strongly affects self-righting outcome by changing mechanical energy budget. Well-coordinated appendage motions (good phase) accumulate more mechanical energy than poorly-coordinated motions (bad phase), thereby better overcoming the potential energy barrier to self-right more successfully. Finally, we demonstrated practical use of the template for predicting a new control strategy to further increase self-righting performance and informing robot design.**

## I. INTRODUCTION

Animals and mobile robots must right themselves when flipped over on the ground [1]–[3]. How likely and how quickly they can self-right is important for survival or continuing operation.

Animals and robots self-right using many strategies. Most robots use a single morphological feature or one type of appendages to self-right (for a review, see [4]), such as adopting an unstable body shape when upside down [5], adding an extruding appendage [6], jumping with elastic energy storage and release [7], and co-opting rounded wings intended for terradynamic streamlining [4], [8]. Few robots have used multiple appendages together to self-right [9]. By contrast, some animals have evolved more complex strategies that combine different body parts to self-right dynamically (for a review, see [10]). This is especially the case when self-righting is strenuous. For example, on slippery surfaces, insects use wings and legs together [10]; turtles with a rigid, flattened shell uses legs and neck together [1].

Coordination between appendages is important in locomotion such as walking [11], running [12], swimming [13], and self-righting [2], for energetic efficiency, recovery from perturbation, maintaining steady-state movement, and faster escaping. Coordinated appendage motions from a central pattern generator also helps simplify the generation of compound movement [14].

Our study focuses on appendage coordination in leg-assisted, winged ground self-righting [10], [15], [16], a strategy observed in the discoid cockroach (Sec. II). In a recent study, we discovered that wing-leg coordination strongly affects the outcome of this strategy [16] (Sec. II). Here, we take the next step in understanding the mechanism by developing a template of leg-assisted, winged self-righting (Sec. III) to model its complex hybrid dynamics (Sec. IV). We used the template to analyze mechanical energy budget during self-righting (Sec. V)—whether the total mechanical energy accumulated (injection by wings and legs minus dissipation) exceeds the potential energy barrier. We discovered that well-coordinated wing opening and leg flailing motions help accumulate more mechanical energy than poorly-coordinated ones, making it easier to overcome the potential energy barrier. In addition, we used the template to devise a new control strategy to coordinate wing and leg motions, which further improved self-righting performance (Sec. VI), and we used template analysis to inform robot design (Sec. VII). Finally, we summarize our contributions and discuss future directions (Sec. VIII).

## II. MODEL SYSTEM & PREVIOUS WORK

When flipped over on a level, flat surface, the discoid cockroach often opens its wings to push against the ground, in an attempt to self-right by a somersault (pitching over the head) [10]. However, because the potential energy barrier in doing so is great (seven times the mechanical energy generated per stride during medium-speed running at 5 body length/s [10]), wings pushing rarely generates sufficient mechanical energy [15] to complete a somersault [10]. While the wings are open and body is pitched up, the animal often flail its legs in the air laterally (Fig. 1A), as well as twist and bend its raised abdomen [10]. These secondary motions perturb the body laterally and facilitate self-righting, by adding kinetic energy [15] and shifting the center of mass. As a result, the animal often eventually self-rights by body rolling after the initial pitching up (Fig. 1B) [10]. For this leg-assisted, winged self-righting strategy, flailing legs is an interval motion which does not change total angular momentum directly. Thus, the animal often makes multiple

Manuscript received: February 24, 2020; Revised May 26, 2020; Accepted July 6, 2020. This paper was recommended for publication by Editor Xinyu Liu upon evaluation of the Reviewers' comments. This work was supported by an Army Research Office Young Investigator award, a Burroughs Wellcome Career Award at the Scientific Interface, and The Johns Hopkins University Whiting School of Engineering start-up funds to C.L.

[1]Qihan Xuan and Chen Li are with the Department of Mechanical Engineering, Johns Hopkins University. (email: [redacted])

wing opening attempts [10] before it accumulates sufficient mechanical energy to overcome the barrier to self-right.

In recent studies, we developed a cockroach-inspired self-righting robot [15] and a multi-body dynamics simulation robot [16] (Fig. 1C; Multimedia Material). We performed simulation experiments to empirically study how wing-leg coordination affects self-righting outcome [16]. The robot has two wings that open symmetrically and a pendulum leg that flails laterally, as well as a curved head that protrudes forward. Its configuration can be defined by wing and leg angles (Fig. 1D). The wings and leg are controlled by motors to oscillate with simple actuation profiles (Fig. 1E). For every wing opening attempt (every cycle of wing oscillation), wing-leg coordination can be measured by the phase φ between wing and leg oscillations (defined in Fig. 1E caption). We varied phase to test whether the robot self-righted. Unless leg oscillation amplitude was very small or very large, for given leg oscillation amplitude, self-righting was always more successful at some phases (e.g., φ = 0%, 50%, and 100%) than others (e.g., φ = 25% and 80%) (Fig. 1F) [16]. Note that in our previous work, random noise was added to the period of wing oscillation during each cycle to change the phase in every wing opening attempt [16]. Here, we focused on understanding of the role of phase during a single attempt, and random noise was not added.

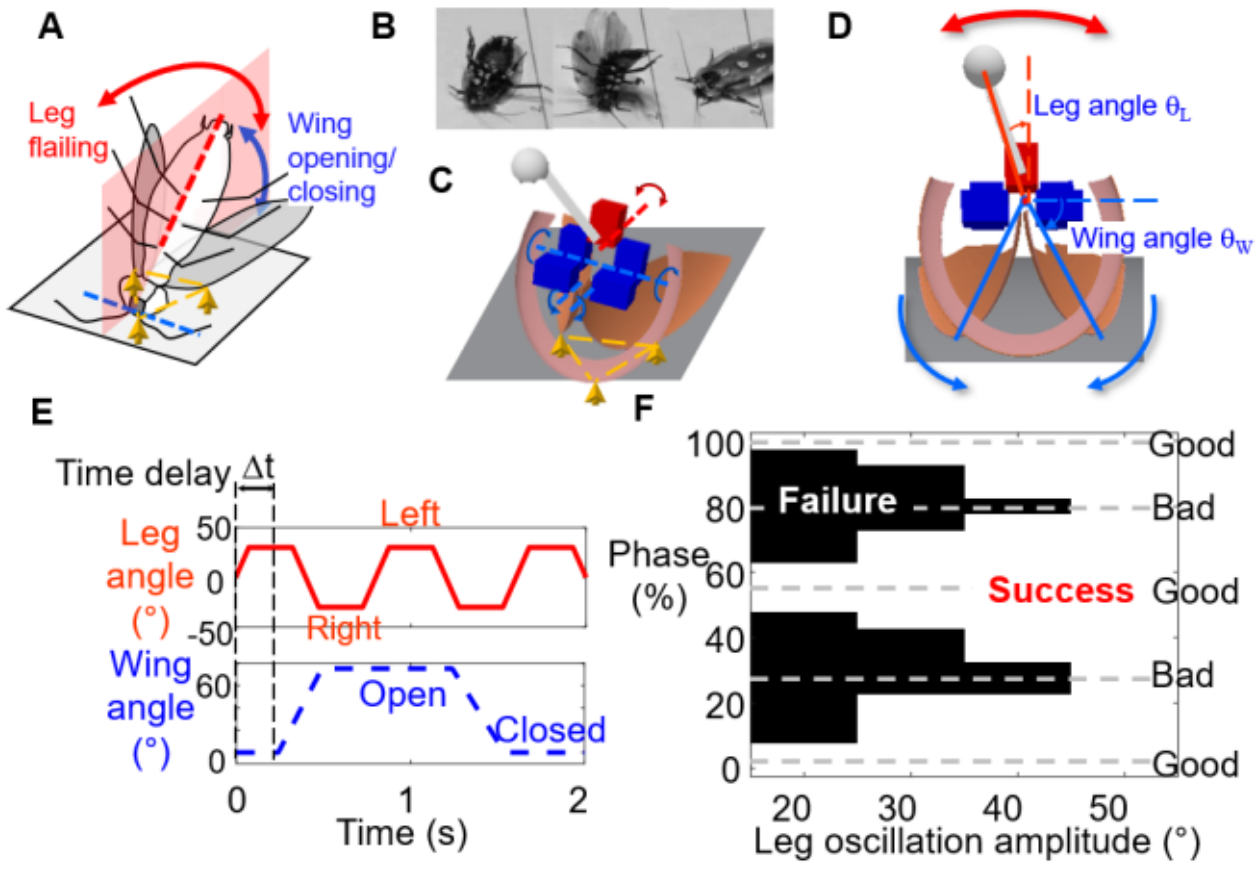

Fig. 1. Leg-assisted, winged self-righting on the ground and previous observations. (A) Schematic of strategy in a discoid cockroach. Yellow arrows and triangle show metastable triangular base of support formed by head and two wings in contact with ground. (B) Snapshots of a cockroach self-righting. (C) Simulation robot with a head, two wings and a pendulum leg. (D) Definition of wing and leg angles. (E) Actuation profiles of wing and leg oscillations. Wing-leg coordination is measured by phase $\varphi = \Delta t/T_{\text{leg}}$, where $\Delta t$ is time delay of start of wing opening from start of the first leg oscillation, and $T_{\text{leg}}$ is leg oscillation period. Wing opening amplitude is the maximal wing opening angle. Leg oscillation amplitude is the maximal leg angle (absolute value) to either side from middle. Leg starts from the middle (leg angle = 0) at time $t = 0$. (F) Self-righting outcome (white: success, black: failure) as a function of phase and leg oscillation amplitude at a given wing opening amplitude (76°), from simulation experiments in previous study [16].

## III. TEMPLATE

Here, to understand the mechanism of how the phase between wing and leg actuation affects self-righting outcome, we developed a template of leg-assisted, winged self-righting (Fig. 2; Multimedia Material).

A template is the simplest dynamical model with the least number of components and degrees of freedom to realize the most fundamental dynamics of a targeted locomotor behavior [17]. Templates have been developed for walking [18], upright [19] and sprawled [20] running, climbing [21], tail-assisted aerial self-righting [22], multi-legged ground self-righting [9], and dynamic large gap traversal [23]. Although they lose details, templates are quantitatively predictive and useful for not only explaining biological phenomena but also predicting design, control, and planning for robots [24].

Our template is novel in that it studies active coordination between different types of appendages. Although appendage coordination is clearly important in the locomotion where more than one appendage is involved, few of the models above systematically studied how appendage coordination affects movement. Arm-leg coordination was studied in walking models, but the arms are passive (not actuated) and active coordination was not studied [11]. The multi-legged ground self-righting model optimally controls the torques of three well-coordinated pairs of legs to maximize thrust and impact kinetic energy recovery [9]; however, such coordination is between appendages of the same type.

For simplicity, our template only models planar dynamics in the transverse plane (Fig. 2), considering that leg-assisted, winged self-righting eventually succeeds by body rolling (see Sec. II). Our template has six simplifying approximations:

(1) The body is a point mass $M$, and the leg mass $m$ is concentrated at its tip.

(2) The wings and leg linkage are massless links only for support and connection.

(3) Wing-ground contact is a point contact. It slides during wing opening but stays fixed when the body rotates around it.

(4) The wings and leg linkage are straight and rigid.

(5) Wing-ground collision is instantaneous and inelastic.

(6) Wing and leg motors start and stop instantaneously.

In our template each contacting wing tip can slide on the ground. From Approximations 5 and 6, wing-ground collision forces and motor torques are infinite (very large compared to robot weight) at the instances of collision and motor starting/stopping.

Although the three-link bipedal walking model [25] appears similar to our template, its targeted locomotor behavior is stable walking as opposed to escaping from a metastable state [15], [16], its ground contact point is unable to slide, and it has different mass properties.

We designed the template to match the robot [15], [16] in geometry and mass distribution (Fig. 2A), and we used the measured coefficient of friction. We also used the same wing and leg actuation profiles as in the simulation robot (Fig. 1E), as shown by their angular velocity profiles (Fig. 2B) [15]. A difference is that motor starting and stopping are instantaneous in the template (Fig. 2B), whereas they take a short time (< 0.04 s) in the robot. We used constant wing opening (300°/s) and closing (250°/s) angular velocities and a constant leg rotation duration (0.15 s), the same as those of the simulation robot. To test the predictive power of the template over a broad range of parameter space, we varied the amplitudes of wing opening (66° to 80° with an increment of 2°) and leg oscillation (20° to 60° with an increment of 10°).

Given the constant leg rotation duration, a larger leg oscillation amplitude corresponds with a larger angular velocity. Wing oscillation periods (2.0 s) and leg oscillation periods (0.8 s) are both constant.

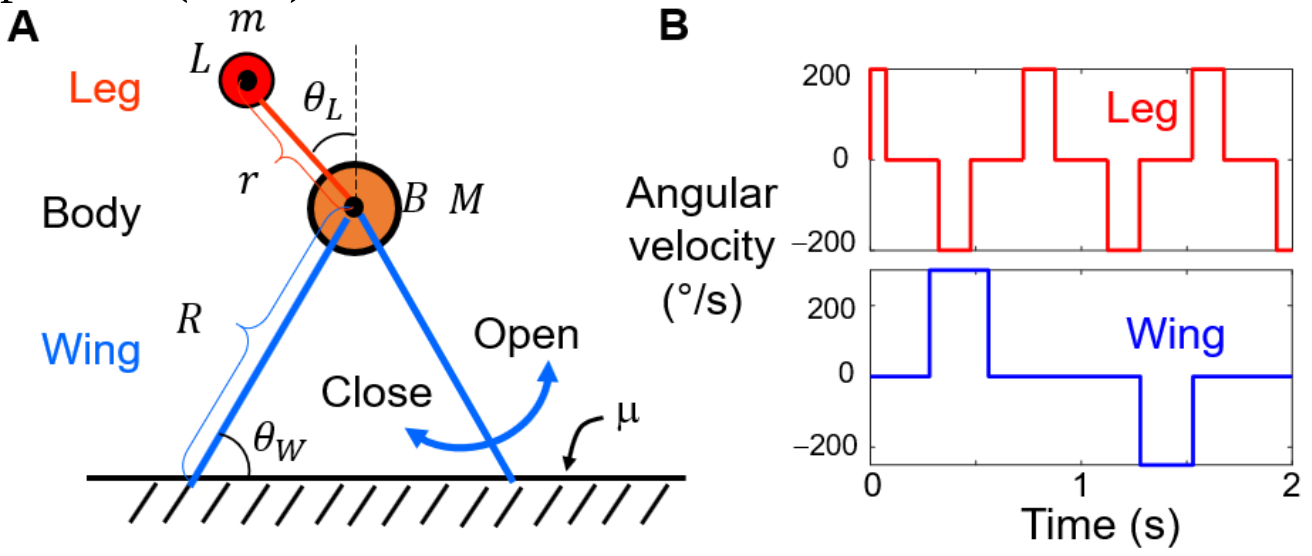

Fig. 2. Template. (A) Schematic of template. Two point masses represent body (orange, B, mass $M$ = 210 g) and leg pendulum mass (red, L, $m$ = 50 g). Three massless links represent wings (blue segments, $R$ = 12 cm) and leg linkage (red segment, $r$ = 7 cm). $\theta_W$ and $\theta_L$ are wing and leg angles. Increasing and decreasing $\theta_W$ correspond with wing opening and closing, respectively. $\mu$ = 1.0 is coefficient of friction between wing and ground measured from experiment using inclined plane method. (B) Actuation profiles of template. Angular velocities of wings (blue) and leg (red) as a function of time.

## IV. DYNAMICS

The template is a hybrid dynamical system [26] because of discontinuity in wing and leg velocities (instantaneous actuation states) and hybrid contact between the wings and ground (instantaneous contact states):

(1) Actuation states. With the prescribed actuation profiles (Fig. 2B), both the wing motor (which opens or closes both wings symmetrically) and leg motor can be in four states—starting, stopping, rotating at a constant angular velocity, and holding an angle. Both wing and leg motor states are prescribed functions of time.

(2) Contact states. Each wing tip can be in four contact states with the ground—no contact, collision, fixed contact, and sliding contact. Given contact states at each time step, we can use the current configuration, position, velocity, and motor states to determine contact states at the next time step.

The actuation and contact states that the system is in affects the forces and external constraints act on it, which often results in different equations of motion.

Some of the states above are continuous motions (e.g., motor rotating or holding an angle in actuation states; fixed or sliding contact and no contact in contact states). Others are instantaneous events (motor starting and stopping in actuation states, collision in contact states). The hybrid dynamics comprises dynamics of continuous motions (e.g., Fig. 3Ai-iii and that of instantaneous events (e.g., Fig. 3Bi-iii).

For continuous motions, we used forces to solve for dynamics. For instantaneous events, we used impulse to solve for the change of velocity via conservation of angular momentum (see Sec. IV, Aiii, Biii).

Below, we describe how we solve the dynamics for six representative cases (Fig. 3), organized into continuous motions (Fig. 3A) and instantaneous events (Fig. 3B). These six cases encompass almost all possible cases of the system, except for when wing motor starts. We write equations in the general form by always including leg rotation angular velocity $\omega_L$; when the leg is holding an angle, we set $\omega_L$ = 0°/s. Initially ($t$ = 0 s), the system is static on the ground with the wings fully closed ($\theta_W$ = 0°) and leg in the middle ($\theta_L$ = 0°). As time elapses, it transitions between different states.

Although the model is 2-D, angular velocity vectors are orthogonal to the transverse plane (i.e., they are along $y$ axis, with $+y$ pointing into the paper). For simplicity, we write equations using 3-D vector format (e.g., $\overrightarrow{\omega_L} = (0, \omega_L, 0)$, $\vec{g} = (0, 0, -g)$, $\overrightarrow{AB} = (x_B, y_B, z_B) - (x_A, y_A, z_A)$, $\overrightarrow{v_B} = (v_{xB}, 0, v_{zB})$).

### A. *Continuous motions*

#### *i. When wings are opening*

When the wings are opening, both wings slide on the ground (Fig. 3Ai). Both wing and leg angular velocities are constant during moving (wing opening or leg rotation) (Fig. 2B):

$$\dot{\theta}_W(t) = \omega_W, \; \dot{\theta}_L(t) = \pm\omega_L \tag{1}$$

where +/– is clockwise/counterclockwise for leg rotation.

Given the angle actuation profiles (Fig. 2B), we can calculate body and leg mass motion in the vertical $z$ direction. Vertical position, velocity, and acceleration of the body are:

$$z_B(t) = R\sin\theta_W(t) \tag{2}$$

$$v_{zB}(t) = \omega_W \, R\cos\theta_W(t) \tag{3}$$

$$a_{zB} = \ddot{z}_B(t) = -R{\omega_W}^2\sin\theta_W(t) \tag{4}$$

Those of the leg mass are:

$$z_L(t) = z_B(t) + r\cos\theta_L(t), \tag{5}$$

$$v_{zL}(t) = v_{zB}(t) - \omega_L \, r\sin\theta_L(t) \tag{6}$$

$$a_{Lz} = \ddot{z}_L(t) = a_{zB} + {\omega_L}^2 r\cos\theta_L(t) \tag{7}$$

From the center of mass theorem, vertical acceleration $a_{zCoM}$ of the center of mass (CoM) is:

$$(M+m)a_{zCoM} = Ma_{zB} + ma_{zL} \tag{8}$$

However, motion in the horizontal $x$ direction cannot be directly calculated from the actuation profiles, and we need to calculate horizontal forces. Below we use "1/2" to refer to the left/right wing. Because of lateral asymmetry in actuation and mass distribution from a single leg rotation, leg torque and gravitational force are asymmetric. Thus, the normal forces $F_1$, $F_2$ and frictional forces $f_1$, $f_2$ at two wing tips are different. This gives the robot center of mass a non-zero lateral velocity.

From Newton's second law, CoM acceleration $a_{xCoM}$ and $a_{zCoM}$ can be solved from:

$$(M+m)a_{xCoM} = f_1 + f_2 \tag{9}$$

$$(M+m)a_{zCoM} = F_1 + F_2 - Mg - mg \tag{10}$$

During wing opening, either both wing tips in contact with the ground slide, or only one slides while the other is fixed. When both slide, frictional forces are:

$$f_1 = \pm\mu F_1 \tag{11}$$

$$f_2 = \pm\mu F_2 \tag{12}$$

where +/– is to the right/left, with frictional force opposite to wing tip horizontal velocity, and μ the coefficient of friction. When only one wing tip slides while the other is fixed, the frictional force on the fixed wing tip does not satisfy Eqns. 11, 12. But because it is fixed, we can calculate the body horizontal velocity directly from wing actuation profile:

$$v_{xB}(t) = \pm\omega_W \, R sin\theta_W(t) \tag{13}$$

where +/– is when the right/left wing tip is fixed.

From the center of mass theorem, horizontal acceleration $a_{xCoM}$ of CoM is given by:

$$(M+m)a_{xCoM} = Ma_{xB} + ma_{xL} \tag{14}$$

Applying Newton's second law in the vertical direction:

$$F_1 + F_2 - (m+M)g = Ma_{zB} + ma_{zL} \tag{15}$$

and about the rotation axis ($y$):

$$\overrightarrow{AC} \times \overrightarrow{F_2} + \overrightarrow{AB} \times M(\vec{g} - \vec{a}_B) + \overrightarrow{AL} \times m(\vec{g} - \vec{a}_L) = 0 \tag{16}$$

With seven equations (Eqns. 9-12, 14-16), we can solve for the seven unknown variables ($F_1$, $F_2$, $f_1$, $f_2$, $a_{xCoM}$, $a_{xB}$, $a_{xL}$). Thus, we can obtain horizontal motion of the system.

*ii. Aerial phase*

During aerial phase, the only external force is gravitational force (Fig. 3Aii, $M\vec{g}$ and $m\vec{g}$). Thus, the CoM moves like a projectile. CoM velocity $\overrightarrow{v_{CoM}}(t)$ can be calculated as a function of time given initial CoM velocity (Sec. IV, Bi). Because the leg is rotating with a constant angular velocity $\overrightarrow{\omega_L}$, the velocities of the leg mass and body follow:

$$\overrightarrow{v_L} - \overrightarrow{v_B} = \overrightarrow{\omega_L} \times \overrightarrow{BL} \tag{17}$$

Further, from the center of mass theorem:

$$(M+m)\overrightarrow{v_{CoM}} = M\overrightarrow{v_B} + m\overrightarrow{v_L} \tag{18}$$

Thus, we can solve these two equations for $\overrightarrow{v_L}$ and $\overrightarrow{v_B}$ during the aerial phase to obtain the projectile dynamics. Because the internal rotational kinetic energy and mechanical energy of CoM are both conserved, the rotating leg motor does no work in the aerial phase.

*iii. When body rotates about a fixed wing tip on ground*

In this case, only gravitational force changes the angular momentum of the system about the contacting wing tip (Fig. 3Aiii). Applying Newton's second law along the rotational direction ($y$ axis), we can solve for body angular velocity $\overrightarrow{\omega_B}(t)$ relative to the contacting wing tip via:

$$(M+m)\vec{g} = \frac{d\vec{L}(t)}{dt} \tag{19}$$

where angular momentum about the contacting wing tip is:

$$\vec{L}(t) = M\overrightarrow{AB} \times \left(\overrightarrow{\omega_B}(t) \times \overrightarrow{AB}\right) + m\overrightarrow{AL} \times \left(\overrightarrow{\omega_B}(t) \times \overrightarrow{AL} + \overrightarrow{\omega_L} \times \overrightarrow{BL}\right) \tag{20}$$

## B. Instantaneous events

*i. When wings stop*

When wing opening stops, there is no external impulse to change the velocities of body and leg mass (Fig. 3Bi). Thus, because of inertia, both wings lift off the ground with a finite upward body velocity (Fig. 3Bi, $v_{zB}$), which can be calculated by applying Eqn. 3 at the instance of wing stopping. The body horizontal velocity remains unchanged at this instance.

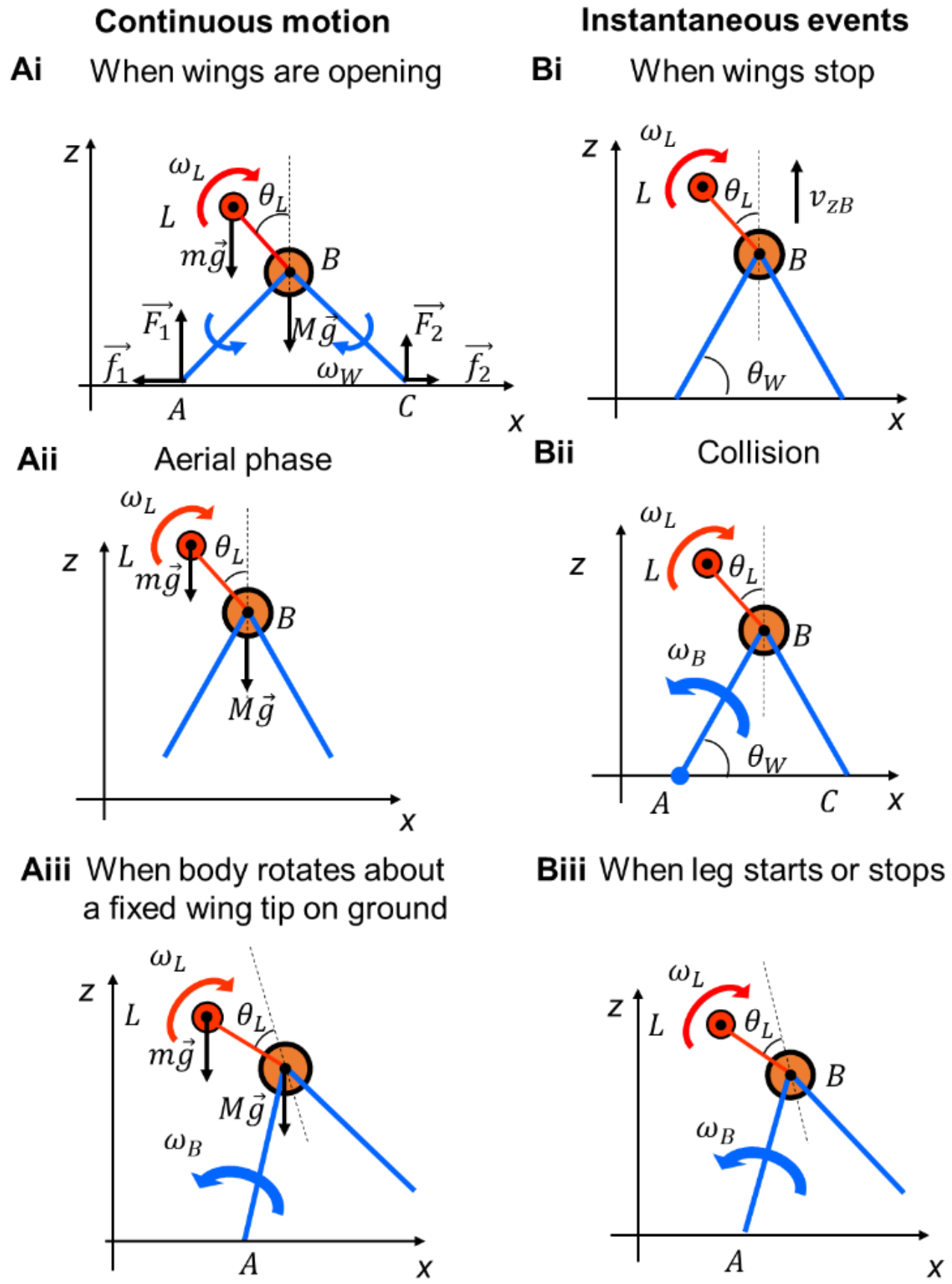

Fig. 3. Hybrid dynamics. We show six representative cases with different actuation and contact states, with definitions of forces ($F_1$, $F_2$, $f_1$, $f_2$, $mg$, $Mg$), wing tips (point A, C), and angular velocity ($\omega_L$, $\omega_B$). (Ai-iii) Continuous motions. (Bi-iii) Instantaneous events.

*ii. Collision*

At collision, two external forces act on the system—gravitational force at the CoM and a collision force at the colliding wing tip (Fig. 3Bii, contact point A). Because the event is instantaneous, the angular impulse generated by the gravitational force can be neglected. In addition, because center of rotation is at the contact point, collision force has no moment arm. Thus, the collision force generates no angular impulse. Therefore, the angular momentum of the system about the instantaneous center of rotation is conserved at collision. We can use this to solve for angular velocity $\overrightarrow{\omega_B}$ of the body:

$$M\overrightarrow{AB} \times \left(\overrightarrow{\omega_B} \times \overrightarrow{AB}\right) + m\overrightarrow{AL} \times \left(\overrightarrow{\omega_B} \times \overrightarrow{AL} + \overrightarrow{\omega_L} \times \overrightarrow{BL}\right) = \vec{L} \tag{21}$$

where the left side is the angular momentum after collision and $\vec{L}$ is the angular momentum before collision, which is solved from the dynamics before collision.

*iii. When leg starts or stops.*

Like during collision (Sec. IV, Bii), when the leg starts or stops, two external forces act on the system—gravitational force at the center of mass and contact force at the contacting wing tip (Fig. 3Biii, contact point A). The contact force is infinite because of infinite leg motor torque. Thus, we can also use Eqn. 21 to solve for the dynamics of leg starting and stopping. For leg stopping, we set $\overrightarrow{\omega_L} = (0, 0, 0)$.

We applied the Euler method to numerically integrate forward in time to calculate the dynamics of the system (both in translation and rotation, with time step $\Delta t = 0.0001$ s). To obtain velocities (both in translation and rotation), we can calculate directly from actuation profiles (Sec. IV), integrate from the acceleration (Sec. IV, Ai-iii), or calculate from the conservation of angular momentum (Sec. IV, Bi-iii).

## V. MECHANICAL ENERGY BUDGET

The robot and the discoid cockroach often struggled for multiple attempts before they could self-right [10], [15], [16]. Although the motors/muscles do positive work when the robot or animal opens the wings and flails the leg(s) (Fig. 1A, C), the mechanical energy injected may not be sufficient to overcome the potential energy barrier. For the robot, this is because mechanical energy is dissipated at collision and motor stopping. To understand how wing-leg coordination affects self-righting outcome, we used the template to calculate how mechanical energy is accumulated from energy injection minus energy dissipation, and we compared cumulative mechanical energy with the potential energy barrier to assess whether self-righting is successful or not.

### *A. Potential energy barrier*

First, we calculated the potential energy barrier of the template. Consider two snapshots during self-righting (Fig. 4A), with CoM at a local minimum (left) and a local maximum (right). A potential energy barrier is the increase in gravitational potential energy from the local minimum to maximum, $E_{\text{barrier}} = (M + m)g\Delta h$. Because of lateral asymmetry in mass distribution from a single leg rotation, rotating to the left and right sides have different potential energy barriers (Fig. 4A). We defined the lower of the two as the potential energy barrier to self-right, as it is the minimal mechanical energy that that must be accumulated. We found that the potential energy barrier decreases with both wing and leg angles (Fig. 4B), consistent with the 3-D robot [15], [16]. For given mechanical energy, the smaller the barrier, the easier it is to self-right.

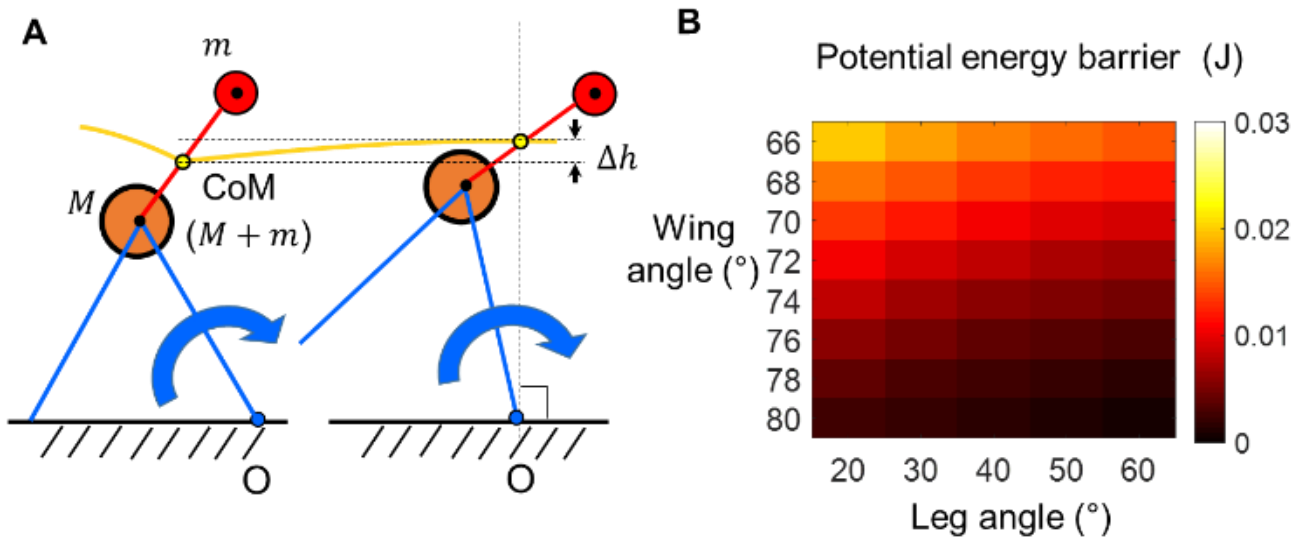

Fig. 4. Potential energy barrier. (A) Two snapshots of template rotating about a fixed wing tip O. Yellow curve is CoM trajectory. $\Delta h$ is CoM height increase between snapshots. (B) Potential energy barrier from template as a function of wing and leg angles.

### *B. Mechanical energy evolution*

Next, we used the template to calculate mechanical energy evolution of the system over the potential energy landscape. We used body roll $\gamma$ to define system orientation (Fig. 5A) and calculated the potential energy landscape over body roll (Fig. 5C-E, blue curve). The system's mechanical energy (Fig. 5C-E, red dot) evolves (red trajectory) over the changing potential energy landscape (blue curve; see Multimedia Material). It is always on or above the potential energy landscape because kinetic energy cannot be negative.

Both wing opening and leg stroke do work to change mechanical energy. In simulation experiments, self-righting always occurs after wing opening stops [16], so we consider subsequent mechanical energy accumulation (Fig. 5C-E).

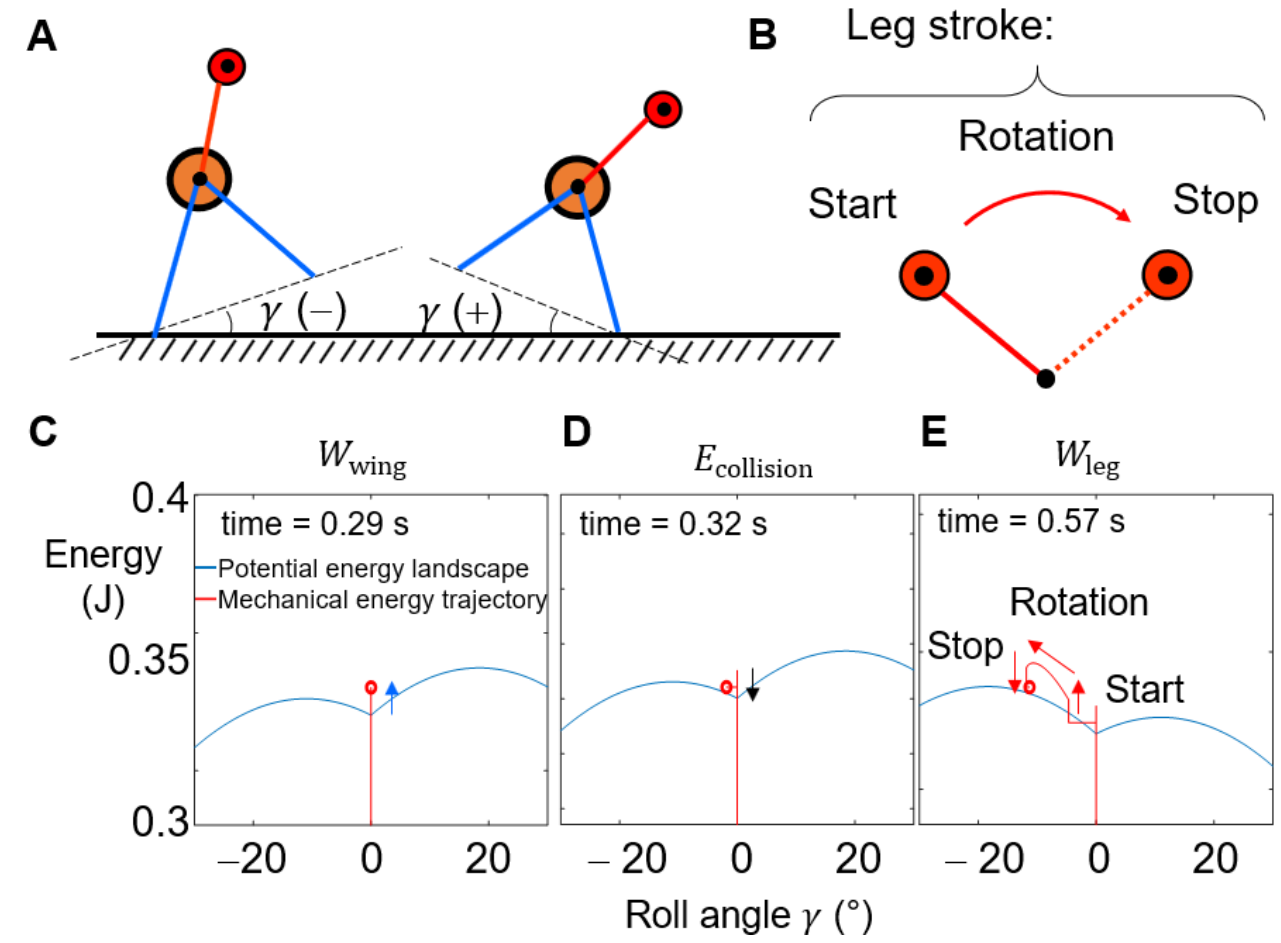

Fig. 5. Mechanical energy evolution of system from template. (A) Definition of roll angle. +/– is clockwise/counterclockwise rolling. (B) Definition of a leg stroke. Stroke direction can be reversed. (C-E) Example snapshots of system mechanical energy (red dot) evolution (red trajectory) over changing potential energy landscape over body roll (blue curve). C: When wing opening stops. D: When one wing collides with ground. E. When leg stops. Red curve is trajectory of system state from the beginning ($t = 0$ s).

We defined wing work $W_{\text{wing}}$ as the work done by the wing motors minus the increase of potential energy and frictional dissipation before wing opening stops. For good phases, $W_{\text{wing}} = E_0 - E_{\text{min}}$, where $E_0$ is the mechanical energy at the instance when wing opening stops and $E_{\text{min}}$ is the local minimum in potential energy landscape. $W_{\text{wing}}$ equals the kinetic energy when wing opening stops (Fig. 5C, blue arrow). For bad phases, $W_{\text{wing}} = E_0 - E_{\text{min}} - W_{\text{leg-before}}$, where $W_{\text{leg-before}}$ is the work done by the leg motor before wing opening stops, which is part of the leg work during a stroke $W_{\text{leg}}$ (defined below). A leg stroke consists of motor starting, rotation, and stopping (Fig. 5B).

After wing opening stops, only the leg motor injects energy into the system. We defined leg work $W_{\text{leg}}$ as the work done by the first leg stroke. In the example shown, both motor starting and rotation inject energy, whereas motor stopping dissipates energy (Fig. 5E, red arrows).

After wing opening stops, there is a brief aerial phase. Leg rotation in the aerial phase does not change mechanical energy (Sec. IV, Aii). After the aerial phase, mechanical energy is dissipated at collision $E_{\text{collision}}$ (Fig. 5D, black arrow). After the first leg stroke, the leg motor is static for 0.25 s (Fig. 2B, top). If there is no second collision, the mechanical energy of the template is conserved. If there is a second collision, it further adds to energy dissipation $E_{\text{collision}}$.

We defined cumulative mechanical energy at the end of the first leg stroke as:

$$E_{cum} = W_{wing} + W_{leg} - E_{collision} \quad (22)$$

The mechanical energy difference between cumulative mechanical energy and potential energy barrier, $E_{\text{cum}} - E_{\text{barrier}}$, equals the excess kinetic energy with which the CoM can rotate in an attempt to overcome the potential energy barrier (Fig. 4, $(M + m)g\Delta h$). Thus, self-righting succeeds if there is a mechanical energy surplus and fails if there is a deficit.

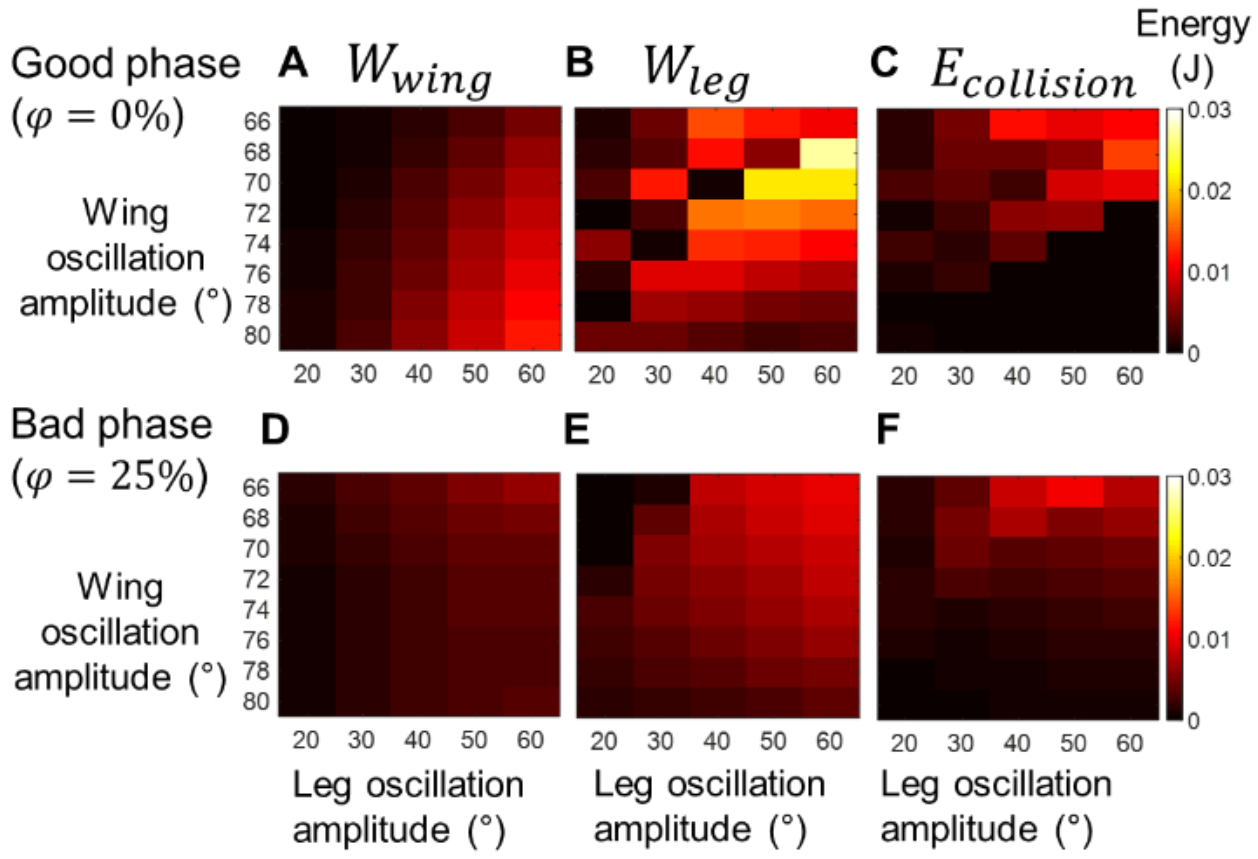

Fig. 6. Mechanical energy contributions for example good (top) and bad (bottom) phases. (A, D) Wing work $W_{\text{wing}}$ (see definition in text), (B, E) work done by the first leg stroke $W_{\text{leg}}$, and (C, F) energy dissipation at collision $E_{\text{collisoin}}$ (absolute value), all as a function of wing opening and leg oscillation amplitudes.

## C. *Effect of phase*

To understand how phase affects system mechanical energy, we calculated contributions to cumulative mechanical energy from wings, leg stroke, and collision (Fig. 6, $W_{\text{wing}}$, $W_{\text{leg}}$, and $E_{\text{collision}}$) for good and bad phases over a broad range of wing opening and leg oscillation amplitudes. Both $W_{\text{wing}}$ and $W_{\text{leg}}$ consist contributions from instantaneous events and continuous motions, whereas $E_{\text{collisoin}}$ is only from instantaneous events. Then, we evaluated mechanical energy budget, $E_{\text{cum}} - E_{\text{barrier}}$, to understand why good phases lead to more successful self-righting than bad phases.

For both good and bad phases, the system has an increasing mechanical energy deficit as wing opening and leg oscillation amplitudes decrease, and it has an increasing surplus as they increase (Fig. 7A, B). Where the template transitions from having an energy surplus to a deficit (Fig. 7A, B, black curve) predicts where it transitions from success to failure (Fig. 7C, D). Clearly, good phases lead to greater mechanical energy surplus and smaller deficit than bad phases over the entire parameter space (Fig. 7A vs. 7B), resulting in successful self-righting over a larger range of wing opening and leg oscillation amplitudes (Fig. 7C vs. 7D). These predictions of self-righting outcome (Fig. 7C, D) are consistent with previous observations in the 3-D simulation robot [16].

Considering that our previous observations were made in a different, 3-D robot [16], to further validate the template, we developed a "2-D" multi-body dynamics simulation robot using Chrono [27] (Fig. 7B, inset; Multimedia Material). We chose Chrono because it is good at handling of complex dynamic systems with large number of rigid bodies that interact through repulsive, frictional contact [27]. The "2-D" robot has the same design as the template except that, to constrain motion within the transverse plane, it has a finite thickness into the plane. We used Young's modulus $E = 1 \times 10^5$ Pa, Poisson's ration $\nu = 0.1$, coefficient of restitution CoR = 0, coefficient of friction $\mu = 1$, and time step t = 0.00001 s.

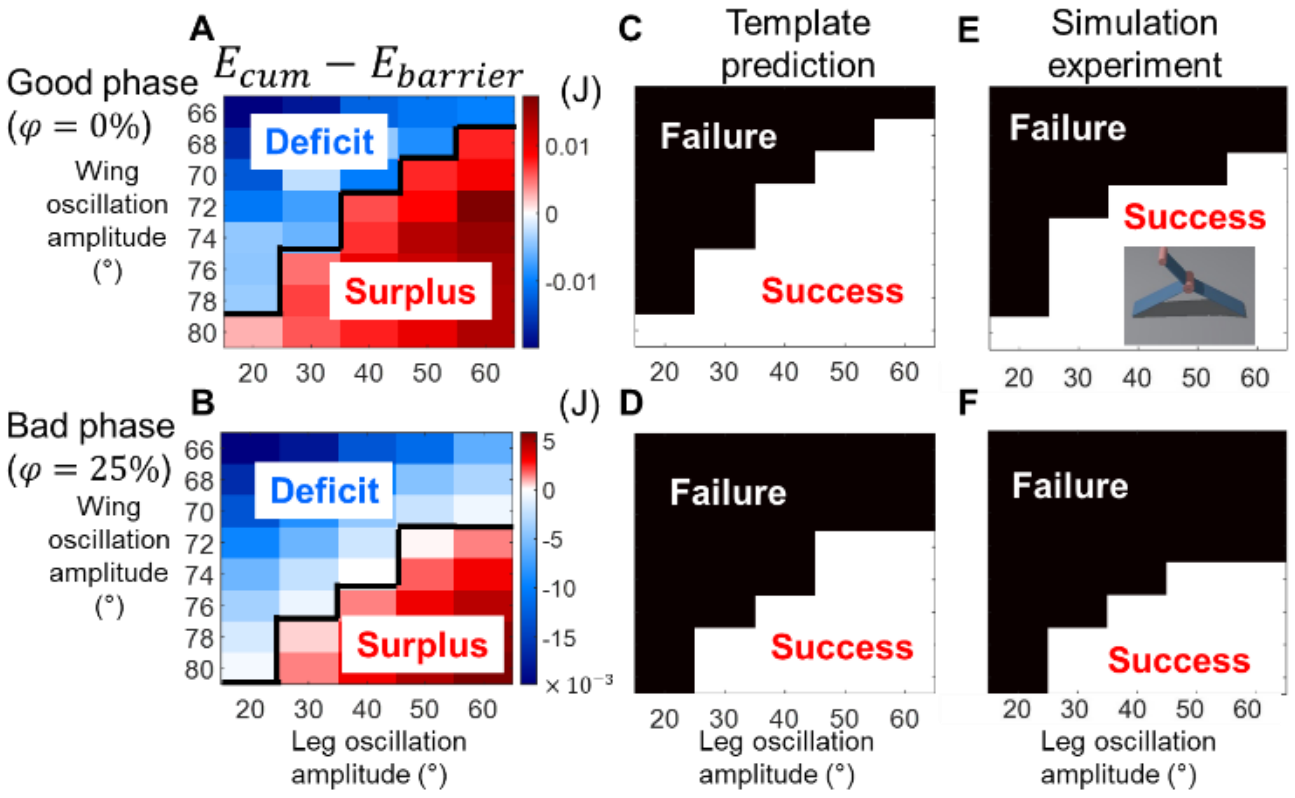

Fig. 7. Mechanical energy budget from template predicts self-righting outcome. (A, B) Mechanical energy budget from template—cumulative mechanical energy minus potential energy barrier. (C, D) Self-right outcome predicted by energy budget from template. (E, F) Self-righting outcome of a "2-D" simulation robot (inset in E). Data are shown as a function of wing opening and leg oscillation amplitudes for good (top) and bad (bottom) phases.

By performing simulation experiments over the same range of parameter space, we found that where the "2-D" robot transitioned from success to failure (Fig. 7E, F) well matched that predicted from mechanical energy budget from the template (Fig. 7A, B), for both good and bad phases. The slight mismatch likely resulted from deviation from model approximations. For example, collision, acceleration, and deceleration are instantaneous in the template, whereas in simulation, these events are fast but still take time. In addition, there is an overlap at contact in simulation that models deformation, but this is not present in the template.

## D. *Effect of leg stroke*

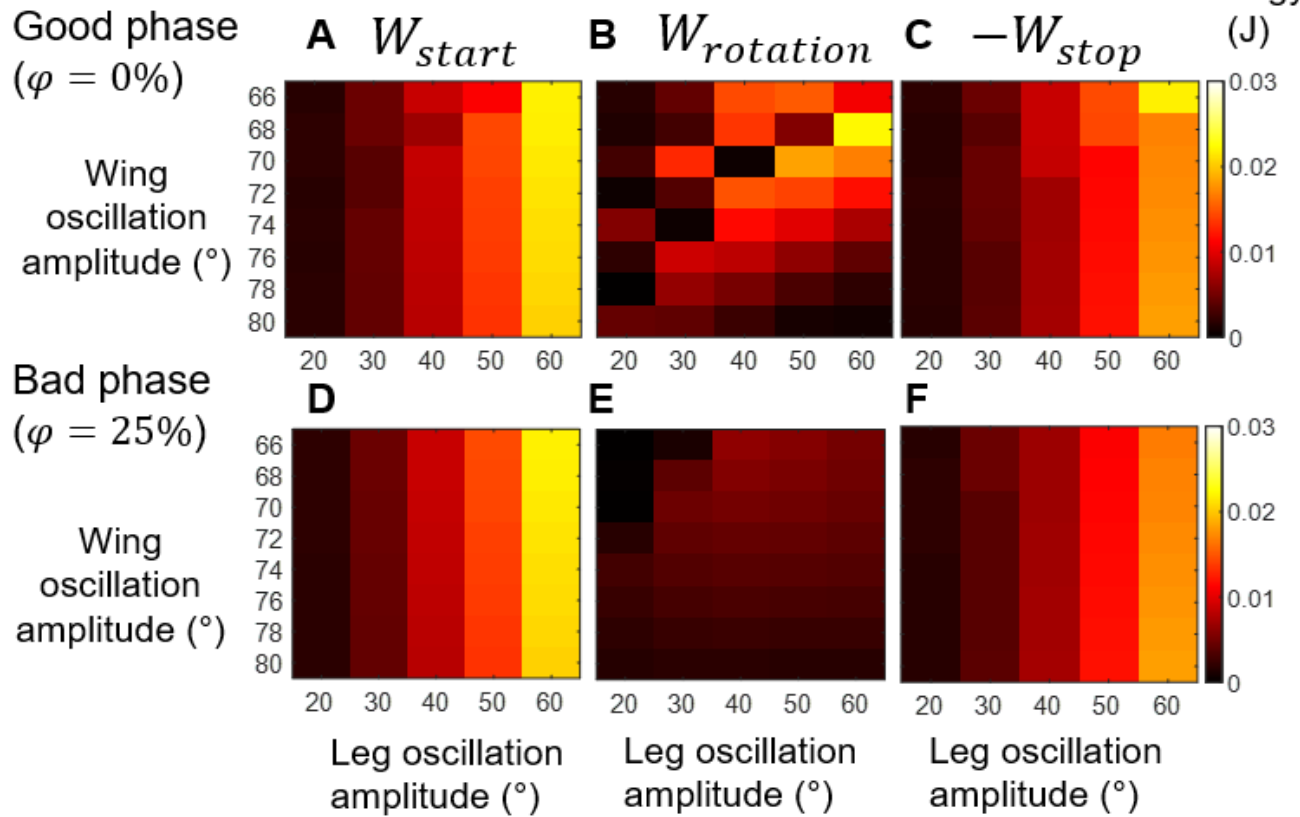

Fig. 8. Contributions to $W_{\text{leg}}$ from leg starting $W_{\text{start}}$, rotation $W_{\text{rotation}}$, and stopping $W_{\text{stop}}$ for example good (top) and bad (bottom) phases.

Our energy calculations also revealed that leg stroke plays a dominant role in self-righting. Although energy contributions from wing opening, leg stroke, and collision depended on phase in complex ways (Fig. 6), work done by leg stroke $W_{leg}$ is the most sensitive to phase (Fig. 6B vs. 6E). In addition, the difference in $W_{leg}$ between good and bad phases (Fig. 6B vs. 6E) is the greatest around the transition boundary in mechanical energy budget (Fig. 7A).

$W_{leg}$ has three contributions (Fig. 8) from leg motor starting, stopping, and rotation (Fig. 5B, E). $W_{start}$ and $W_{stop}$ are from instantaneous events, insensitive to phase (Fig. 8A, C vs. D, F), and nearly cancel each other (Fig. 8A, D vs. C, F). By contrast, $W_{rotation}$ depends sensitively on phase (Fig. 8B vs. E).

## VI. TEMPLATE-PREDICTED CONTROL

Because the template describes system dynamics fully and was validated against simulation experiments, we can use it for predictive robot control. To demonstrate this, we used the template to predict a new control strategy to further increase self-righting performance.

Because $W_{rotation}$ plays a dominant role in mechanical energy budget (Sec. V, D), our new control strategy maximizes work done by the leg during rotation:

$$W_{rotation} = \int_{-\theta_L}^{\theta_L} \tau_{leg}\, d\theta \qquad (23)$$

where $\tau_{leg}$ is leg motor torque. We calculated leg torque as a function of time:

$$\tau_{leg}(t) = \frac{dW_{leg}(t)}{d\theta(t)} = \frac{dW_{leg}(t)}{dt}\frac{dt}{d\theta(t)} = \frac{dW_{leg}(t)}{\omega dt} \qquad (24)$$

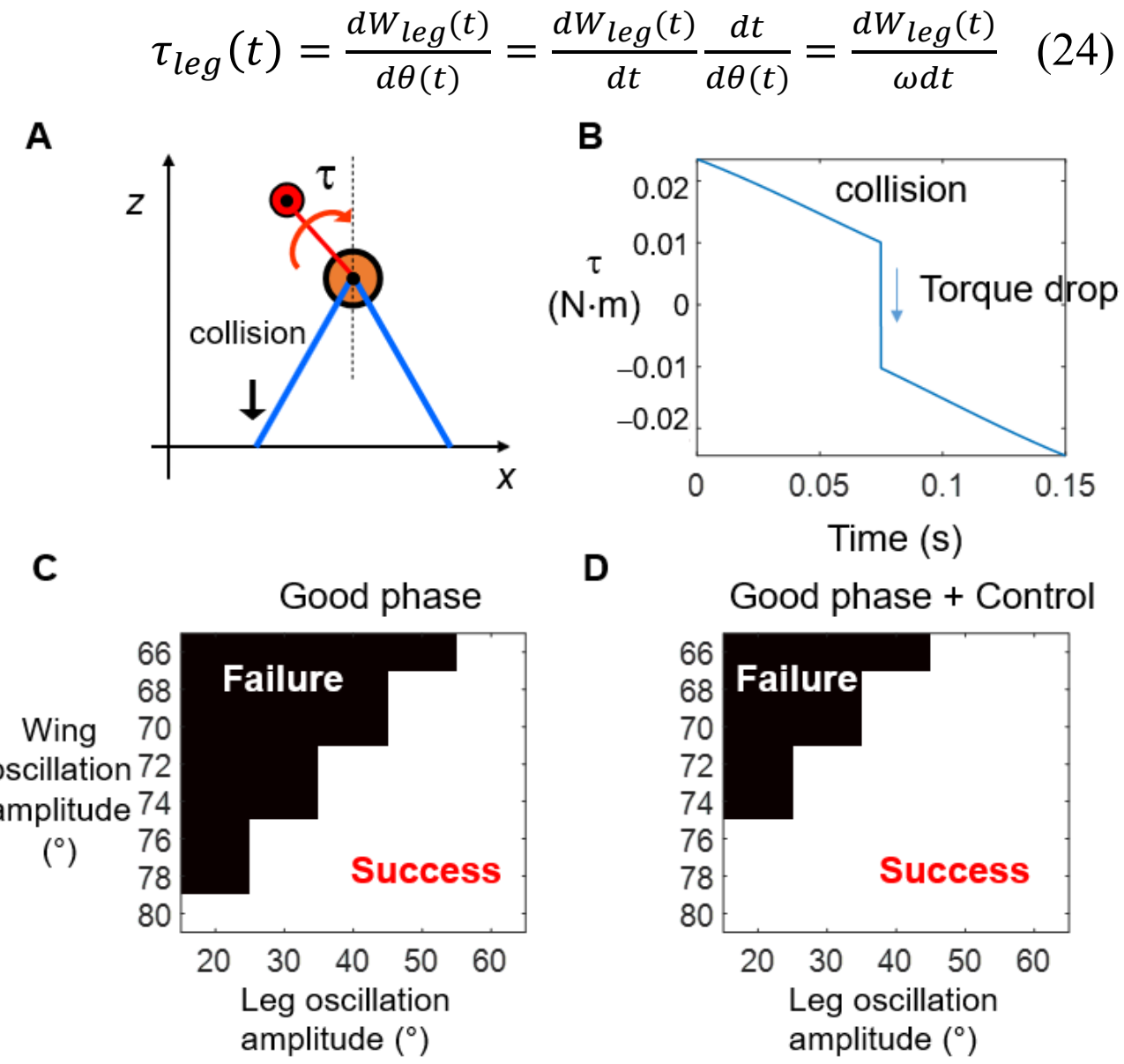

Fig. 9. Template-predicted control. (A) A collision with ground occurs during a leg stroke. τ is leg motor torque. (B) Leg motor torque τ as a function of time during a leg stroke. Torque drops at collision. (C, D) Self-righting outcome predicted from template with (C) and without (D) control using a good phase (φ = 50%), both as a function of wing opening and leg oscillation amplitudes.

Our calculation revealed that, when a wing collides with the ground (Fig. 9A), leg torque drops instantaneously (Fig. 9B). From Eqn. (23), $W_{rotation}$ is maximized when $\tau_{leg}$ is maximized, because integration interval is fixed. Thus, our control strategy maximized $W_{rotation}$ by maximally delaying collision during leg stroke, or avoiding collision altogether.

To achieve this, we start the leg motor immediately after collision, so that the leg stroke does positive work for a longer time before the next collision. This is considering that the time interval between two collisions is nearly unaffected by leg motor starting, because CoM angular momentum relative to the contacting wing tip stays unchanged during leg motor starting (Sec. IV, Biii).

We used the template to test our control strategy. Indeed, self-righting performance was further improved (Fig. 9D) beyond achievable by using good phase alone (Fig. 9C).

## VII. TEMPLATE-INFORMED DESIGN

Next, we use the template to analyze how system parameters affect cumulative mechanical energy and the potential energy barrier to inform robot design.

First, the potential energy barrier is a function of six system parameters (wing and leg angles, body and leg pendulum masses, and body and leg lengths), i.e., $E_{barrier} = E_{barrier}(\theta_W, \theta_L, M, m, R, r)$. If $E_{barrier}$ is monotonic to a parameter, one can increase or decrease this parameter to reduce $E_{barrier}$ for easier self-righting. Thus, we calculated the partial derivative of $E_{barrier}$ over each parameter. If the partial derivative is always $\geq 0$ or $\leq 0$, $E_{barrier}$ increases or decreases monotonically with this parameter regardless of other parameters. We randomly sampled 100 initial conditions and numerically searched for the maximum and minimum of each partial derivative, with the constraint that the projection of CoM on the ground is between two wing tips.

We found that $E_{barrier}$ decreases monotonically with wing angle $\theta_W$ and leg length $r$ and increases monotonically with body mass $M$ and wing length $R$, regardless of other parameters. However, $E_{barrier}$ does not always increase or decrease monotonically with leg angle $\theta_L$ and leg pendulum mass $m$ regardless of other parameters.

Second, cumulative mechanical energy depends on nine system parameters, including wing opening speed $\omega_W$, leg rotation speed $\omega_L$, duration of leg rotation, and the six above. However, because cumulative mechanical energy does not have an analytical form because of the hybrid dynamics, we have not systematically studied how it depends on parameters. Considering that leg velocity is typically much larger than body velocity, kinetic energy of the leg pendulum is:

$$E_{kL} \sim \frac{1}{2} m r^2 {\omega_L}^2 \qquad (25)$$

Thus, increasing $m$, $r$, $\omega_L$ will increase cumulative mechanical energy. In addition, increasing $\theta_W$ will reduce collision dissipation by reducing the moment arm of collision force and thus increase cumulative mechanical energy.

Based on the above, having wings that can open by a larger angle and longer legs that flail faster will likely make a good design for both robots and animals for leg-assisted, winged self-righting. These insights into appendage design added to our knowledge of how body geometry (e.g. turtle shells [1], body aspect ratios [10]) affects ground self-righting.

## VIII. SUMMARY & FUTURE WORK

We developed a template of leg-assisted, winged ground self-righting and used it to elucidate why wing-leg coordination affected self-righting outcome. With good coordination, the system better accumulates mechanical energy to overcome the potential energy barrier and thus self-rights more successfully. Besides providing mechanistic understanding of this complex hybrid dynamical system, we demonstrated practical usefulness of the template by predicting a new control strategy that further improves self-righting performance. We also used the template to examine how system parameters may affect cumulative mechanical energy and the potential energy barrier to inform robot design. More broadly, our study highlighted the importance of appendage coordination for appropriately cumulating mechanical energy to overcome potential energy barriers in strenuous locomotor transitions. An energy landscape approach to locomotor transitions [28] will facilitate this progress [10], [15], [29]–[31].

Our template opens an avenue for further studies of leg-assisted, winged ground self-righting. Morphological and actuation details can be added to create models anchored into the template [17], [32] for studying more complex self-righting strategies. For example, we can use two ellipses with distributed mass to model the body with bending abdomen observed in the animal [10] and add multiple oscillating pendulum legs to study how wings, legs, and abdomen coordinate together. We can also revise rigid wing links into wing springs to study how wing deformation affects energy cumulation, as the thin wings can deform under load. Finally, it will be intriguing to use the template to study how animals and robots can sense their state and interaction with the ground to better coordinate appendages to self-right.

## ACKNOWLEDGMENTS

We thank Ratan Othayoth, Tom Libby, Shai Revzen, Simon Sponberg, Yifeng Zhang, and Yaqing Wang for discussion and three anonymous reviewers for suggestions.

## REFERENCES

[1] G. Domokos and P. L. Várkonyi, "Geometry and self-righting of turtles," *Proc. R. Soc. B Biol. Sci.*, vol. 275, no. 1630, pp. 11–17, 2008.

[2] L. Frantsevich, "Righting kinematics in beetles (Insecta: Coleoptera)," *Arthropod Struct. Dev.*, vol. 33, no. 3, pp. 221–235, 2004.

[3] E. Guizzo and E. Ackerman, "The hard lessons of DARPA's robotics challenge [News]," *IEEE Spectr.*, vol. 52, no. 8, pp. 11–13, 2015.

[4] C. Li, C. C. Kessens, R. S. Fearing, and R. J. Full, "Mechanical principles of dynamic terrestrial self-righting using wings," *Adv. Robot.*, vol. 31, no. 17, pp. 881–900, 2017.

[5] M. Kovač, M. Schlegel, J. C. Zufferey, and D. Floreano, "Steerable miniature jumping robot," *Autonomous Robots*, vol. 28, no. 3. pp. 295–306, 2010.

[6] C. S. Casarez and R. S. Fearing, "Dynamic terrestrial self-righting with a minimal tail," *IEEE Int. Conf. Intell. Robot. Syst.*, vol. 2017-Septe, pp. 314–321, 2017.

[7] P. Fiorini and J. Burdick, "The development of hopping capabilities for small robots," *Auton. Robots*, vol. 14, no. 2–3, pp. 239–254, 2003.

[8] C. Li, A. O. Pullin, D. W. Haldane, H. K. Lam, R. S. Fearing, and R. J. Full, "Terradynamically streamlined shapes in animals and robots enhance traversability through densely cluttered terrain.," *Bioinspir. Biomim.*, vol. 10, no. 4, p. 046003, 2015.

[9] U. Saranli, A. A. Rizzi, and D. E. Koditschek, "Model-based dynamic self-righting maneuvers for a hexapedal robot," *Int. J. Rob. Res.*, vol. 23, no. 9, pp. 903–918, 2004.

[10] C. Li, T. Wöhrl, H. K. Lam, and R. J. Full, "Cockroaches use diverse strategies to self-right on the ground," *J. Exp. Biol.*, vol. 222, no. 15, p. jeb186080, 2019.

[11] S. H. Collins, P. G. Adamczyk, and A. D. Kuo, "Dynamic arm swinging in human walking," *Proc. R. Soc. B Biol. Sci.*, vol. 276, no. 1673, pp. 3679–3688, 2009.

[12] J. F. Seay, J. M. Haddad, R. E. A. van Emmerik, and J. Hamill, "Coordination variability around the walk to run transition during human locomotion," *Motor Control*, vol. 10, no. 2, pp. 178–196, 2006.

[13] B. Kwak and J. Bae, "Design of hair-like appendages and comparative analysis on their coordination toward steady and efficient swimming," *Bioinspiration and Biomimetics*, vol. 12, no. 3, 2017.

[14] M. C. Tresch, P. Saltiel, A. D'Avella, and E. Bizzi, "Coordination and localization in spinal motor systems," *Brain Res. Rev.*, vol. 40, no. 1–3, pp. 66–79, 2002.

[15] R. Othayoth, Q. Xuan, and C. Li, "Leg vibrations help cockroaches self-right using wings," in *Integrative and Comparative Biology*, vol. 57, pp. E370, 2017.

[16] Q. Xuan and C. Li, "In silico experiments reveal the importance of randomness of motions in cockroach's winged self-righting," in *Integrative and Comparative Biology*, vol. 59, pp. E255, 2019.

[17] R. J. Full and D. E. Koditschek, "Templates and anchors: Neuromechanical hypotheses of legged locomotion on land," *J. Exp. Biol.*, vol. 202, no. 23, pp. 3325–3332, 1999.

[18] G. A. Cavagna, N. C. Heglund, and C. R. Taylor, "Mechanical work in terrestrial locomotion: two basic mechanisms for minimizing energy expenditure," *Am. J. Physiol. - Regul. Integr. Comp. Physiol.*, vol. 2, no. 3, 1977.

[19] R. Blickhan and R. J. Full, "Similarity in multilegged locomotion: Bouncing like a monopode," *J. Comp. Physiol. A*, vol. 173, no. 5, pp. 509–517, 1993.

[20] J. Schmitt and P. Holmes, "Mechanical models for insect locomotion: Dynamics and stability in the horizontal plane I. Theory," *Biol. Cybern.*, vol. 83, no. 6, pp. 501–515, 2000.

[21] D. I. Goldman, T. S. Chen, D. M. Dudek, and R. J. Full, "Dynamics of rapid vertical climbing in cockroaches reveals a template.," *J. Exp. Biol.*, vol. 209, no. Pt 15, pp. 2990–3000, Aug. 2006.

[22] T. Libby, A. M. Johnson, E. Chang-Siu, R. J. Full, and D. E. Koditschek, "Comparative Design, Scaling, and Control of Appendages for Inertial Reorientation," *IEEE Trans. Robot.*, vol. 32, no. 6, pp. 1380–1398, 2016.

[23] S. W. Gart, C. Yan, R. Othayoth, Z. Ren, and C. Li, "Dynamic traversal of large gaps by insects and legged robots reveals a template," *Bioinspiration and Biomimetics*, vol. 13, no. 2, 2018.

[24] P. Holmes, R. J. Full, D. Koditschek, and J. Guckenheimer, "The dynamics of legged locomotion: Models, analyses, and challenges," *SIAM Rev.*, vol. 48, no. 2, pp. 207–304, 2006.

[25] J. W. Grizzle, G. Abba, and F. Plestan, "Asymptotically stable walking for biped robots: Analysis via systems with impulse effects," *IEEE Trans. Automat. Contr.*, vol. 46, no. 1, pp. 51–64, 2001.

[26] P. E. Caines, "An introduction to hybrid dynamical systems," *Automatica*, vol. 38, no. 3, pp. 566–567, 2002.

[27] A. Tasora *et al.*, *Chrono: An Open Source Multi-physics Dynamics Engine*, vol. 9611. 2016.

[28] R. Othayoth, G. Thoms, and C. Li, "An energy landscape approach to locomotor transitions in complex 3-D terrain," *Proc. Natl. Acad. Sci. U. S. A.*, vol. 117, no. 26, pp. 14987-14995, 2020.

[29] S. W. Gart and C. Li, "Body-terrain interaction affects large bump traversal of insects and legged robots," *Bioinspiration and Biomimetics*, vol. 13, no. 2, pp. 026005, 2018.

[30] Y. Wang, R. S. Othayoth Mullankandy, and C. Li, "Active adjustments help cockroaches traverse obstacles by lowering potential energy barrier," *Bull. Am. Phys. Soc.*, pp. S22--003, 2020.

[31] Y. Han, Y. Wang, and C. Li, "Body shape helps legged robots climb and turn in complex 3-D terrains," in *Bulletin of the American Physical Society 62*, pp. Y12--002, 2017.

[32] J. Lee, S. N. Sponberg, O. Y. Loh, A. G. Lamperski, R. J. Full, and N. J. Cowan, "Templates and anchors for antenna-based wall following in cockroaches and robots," *IEEE Trans. Robot.*, vol. 24, no. 1, pp. 130–143, 2008.

Multimedia Materia: The video has three sections:

1. An example video for the 3-D simulation robot self-righting.
2. The difference in self-righting between bad and good phases. We kept other parameters the same and only changed the phase. The top videos show the template motion and the bottom videos show the corresponding mechanical energy evolution on potential energy landscape.
3. Example videos of "2-D" multi-body dynamics simulation robot with bad and good phases.